\documentclass[twocolumn,aps,pra,10pt,showpacs]{revtex4-2}
\usepackage{bm}
\usepackage{amsfonts}
\usepackage{amssymb}
\usepackage{amsmath}
\usepackage{graphicx}

\begin{document}
\title{Comment on "Possibility of small electron states"}
\author{Iwo Bialynicki-Birula}\email{birula@cft.edu.pl}
\affiliation{Center for Theoretical Physics, Polish Academy of Sciences\\
Aleja Lotnik\'ow 32/46, 02-668 Warsaw, Poland}
\author{Zofia Bialynicka-Birula}
\affiliation{Institute of Physics, Polish Academy of Sciences\\
Aleja Lotnik\'ow 32/46, 02-668 Warsaw, Poland}

\begin{abstract}
It is shown that the interpretation of the electron wave function as a classical field is untenable because the so called energy-density defined in \cite{seb} takes on negative values in some regions. The claim that the velocity of the electron never exceeds the speed of light is also invalid. The velocity, as defined by the author, becomes even infinite at some points.
\end{abstract}

\maketitle

In the recent paper \cite{seb} the author studied the problem of the size of the relativistic electron wave functions obeying the Dirac equation. We argue that the interpretation of the electron wave functions as if they were describing the behavior of some kind of a classical physical medium is untenable because it leads to paradoxes and contradictions. We show by the explicit calculation that the notions of the energy density and of the velocity of the electron wave packet used by the author are marred by serious flaws.

In the first part of this Comment we study in some detail the properties of the Gaussian solutions of the Dirac equation employed in \cite{seb}. The probability density for the Gaussian wave packet has the form of an expanding spherical ball whose average value of $\bm r^2$ is a quadratic function of time,
\begin{align}
\langle{\bm r}^2\rangle=A+B\,t^2.
\end{align}
The dependence on $t^2$ characterizes the spreading of all freely evolving wave packets in relativistic and also in the nonrelativistic theory. The quadratic dependence on $t$ is a simple result of the fact that for freely moving particle $\bm r$ is proportional to $t$. The same result is valid, of course, for freely expanding ball of gas. One is tempted, therefore, to view the probability distribution $\psi^\dagger\psi$ for the Dirac particle as representing some kind of a moving substance. In our opinion, the author of \cite{seb} has succumbed to this temptation. This issue becomes important when it comes to the problem of the local energy distribution and the velocity. The author refers to the energy distribution given by Eq.~(18) as the {\em energy density} but this term is highly misleading. The so called density of the energy $\rho^{\mathcal{E}}$ discussed in \cite{seb} {\em does not satisfy} the requirement of positivity. Therefore, one should avoid the use of the term density or, at least, replace it with the term {\em quasi-density}.

In order to underscore this point, we will calculate explicitly the quasi-density of the energy for the Gaussian state $f(p)=\exp(-l^2p^2)$ considered in \cite{seb}. The calculation of the Dirac bispinor is simplified if one uses the prescription described in \cite{bb} which connects the solutions of the Dirac equation for the bispinor with the derivatives of the solutions of the Klein-Gordon equation. This will enable us to perform the integration over the angles and obtain the solutions in terms of spherical waves instead of plane waves. The solution of the Klein-Gordon with positive frequency for the Gaussian wave packet is given by the following integral $(\hbar=1,\,c=1)$,
\begin{align}\label{kg}
\phi(x,y,z,t)=\int_0^\infty\!\!\!dp\,j_0(p r)
e^{-i E_p t}e^{-l^2p^2},
\end{align}
where $r=\sqrt{x^2+y^2+z^2},\,E_p=\sqrt{m^2+p^2}$, and $j_0(p r)$ is the spherical Bessel function. The corresponding solution of the Dirac equation with spin along the $z$-axis is obtained by the differentiation with respect to space-time coordinates,
\begin{align}\label{kg2d}
\psi(x,y,z,t)=\left[\begin{array}{c}(i\partial_t+m)\phi\\
0\\-i\partial_z\phi\\
-i(\partial_x+i\partial_y)\phi\end{array}\right].
\end{align}
We used in this formula the standard Dirac representation, instead of the Weyl representation of the matrices used in \cite{bb}. The derivatives of $\phi$ with respect to the space-time variables can be evaluated under the integral sign and we arrive at the following representation of the Dirac bispinor $\psi$ in the form of a one-dimensional integral,
\begin{widetext}
\begin{align}\label{psi}
\psi(x,y,z,t)=\int_0^\infty\!\!\!dp\,u(x,y,z)
e^{-i E_p t}e^{-l^2p^2}=\int_0^\infty\!\!\!dp\,
\left[\begin{array}{c}(E_p+m)j_0(pr)\\
0\\i\,p\,z\,j_1(pr)/r\\
ip(x+iy)j_1(pr)/r\end{array}\right]e^{-i E_p t}e^{-l^2p^2}.
\end{align}
\end{widetext}
The bispinor $\psi(x,y,z,t)$ is a solution of the Dirac equation because the bispinor $u(x,y,z)$ obeys the equation,
\begin{align}\label{eig}
\left(-i(\alpha_x\partial_x+\alpha_y\partial_y+
\alpha_z\partial_z)+m\beta-E_p\right)u(x,y,z)=0.
\end{align}
The formula for $\rho^{\mathcal{E}}$ given in \cite{seb} by Eq.~(18),
\begin{align}\label{rho}
\rho^{\mathcal{E}}=\frac{i}{2}(\psi^\dagger\partial_t\psi
-\partial_t\psi^\dagger\psi),
\end{align}
coincides with the expression for the time component of the energy-momentum tensor for the Dirac wave function \cite{wp}. It contains $\psi$ and also the time derivative of $\psi$. The expression for the time derivative is obtained by placing the factor $E_p$ inside the integral in (\ref{psi}).

\begin{figure}[t]
\begin{center}
\includegraphics[width=0.386\textwidth,
height=0.18\textheight]{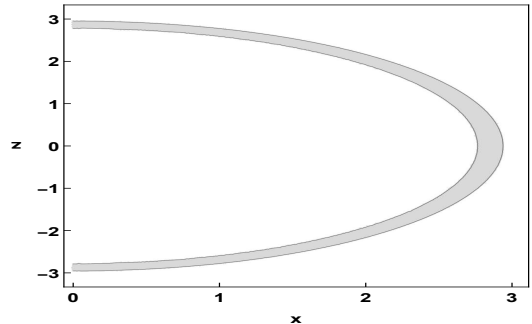}
\caption{The quasi-energy distribution for the Gaussian wave packet with the following set of parameters: $l=0.1,\,t=2.5$. The variables $l,\,x,\,z$ and $t$ are measured in natural units $\hbar/mc$ and $\hbar/mc^2$, respectively.}\label{fig1}
\end{center}
\end{figure}

In our calculation the expression for $\rho^{\mathcal{E}}$ contains the products of two integrals over $p$. These integrals cannot be evaluated analytically but the numerical integration of one-dimensional integrals does not present any problems. The resulting expression for $\rho^{\mathcal{E}}$ has the cylindrical symmetry around the $z$ axis. Therefore, to visualize this function it is sufficient to show the cut by the $y$ plane. The results are presented in Fig.~\ref{fig1}. In this figure we marked the contour lines corresponding to $\rho^{\mathcal{E}}(\bm{r},t)=0$ in the half of the $y=0$ plane. The gray stripe between the zero contour lines corresponds to {\em negative values} of $\rho^{\mathcal{E}}$. In the full three-dimensional plot the gray stripe becomes a thickened ellipsoidal shell obtained by rotation. Of course, the total energy, i.e. the integral over all space is positive, because it is the quantum-mechanical expectation value of the energy operator and the solutions of the Dirac equation studied here are built solely from positive energy waves. This fact also explains why the regions with negative values are so scarce.

The vanishing of $\rho^{\mathcal{E}}$ contradicts the claim by the author that the speed of light is not exceeded. Since $\rho^{\mathcal{E}}$ and $\vec{G}$ are defined by Eqs.~(18) and (19), Eq.~(17) is de facto the definition of the velocity $\vec v$,
\begin{align}\label{vel}
\vec{v}=\frac{\vec{G}}{\rho^{\mathcal{E}}}.
\end{align}
We checked that $\vec{G}$ does not vanish on the surfaces where $\rho^{\mathcal{E}}=0$.  Thus, it follows from (\ref{vel}) that the velocity dramatically exceeds the speed of light; it is infinite. It is worth mentioning here that there exist a sensible definition of the velocity $\vec{v}_D$ for the Dirac particle built from the current vector and the charge density,
\begin{align}\label{veld}
\vec{v}_D=\frac{\vec{j}}{j_0}
=\frac{\psi^\dagger\vec{\alpha}\psi}{\psi^\dagger\psi},
\end{align}
which does not exceed the speed of light for all bispinors $\psi$.

\begin{figure}[b]
\begin{center}
\includegraphics[width=0.36\textwidth,
height=0.18\textheight]{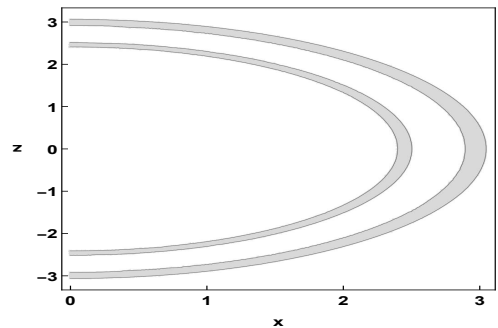}
\caption{The quasi-energy distribution for the Gaussian wave packet with the additional prefactor $(p^2-q^2m^2)$ in the integrand. The parameters are the same as in Fig.~\ref{fig1} and $q=7$.}\label{fig2}
\end{center}
\end{figure}

The appearance of regions with negative values of $\rho^{\mathcal{E}}$ is not a property of the Gaussian solution but it is a common feature. For example, by placing an additional factor $(p^2-q^2 m^2)$ under the integral (\ref{psi}) we obtain not one but two regions of negative values shown in Fig.~\ref{fig2}.

The presence of the regions where $\rho^{\mathcal{E}}$ becomes negative can be proven without using a specific model. Let us consider first a monochromatic solution of the Dirac equation with the energy $E$. In this case we have a positive function  $\rho^{\mathcal{E}}=E|\psi|^2$. The situation changes when we have a sum of two monochromatic solutions $\psi_1+\psi_2$ with the energies $E_1$ and $E_2$. The expression for $\rho^{\mathcal{E}}$ now is,
\begin{align}\label{rho1}
\rho^{\mathcal{E}}=\frac{1}{2}\left(E_1|\psi_1|^2+E_1\psi_2^\dagger\psi_1+
E_2\psi_1^\dagger\psi_2+E_2|\psi_2|^2+c.c.\right).
\end{align}
Let us rescale the first bispinor as follows $\psi_1=\chi m/E_1$, which gives
\begin{align}\label{rho2}
\rho^{\mathcal{E}}=\frac{1}{2}\left(\frac{m|\chi|^2}{E_1}+
m^2\psi_2^\dagger\chi+\frac{mE_2\chi^\dagger\psi_2}{E_1}
+E_2|\psi_2|^2+c.c.\right).
\end{align}
In the limit, when $E_1\to\infty$ or $m/E_1\to 0$, we obtain,
\begin{align}\label{rho3}
\rho^{\mathcal{E}}\to(m\psi_2^\dagger\chi+
E_2|\psi_2|^2+c.c.)/2.
\end{align}
Since the strength and the sign of the bispinor $\chi$ is at our disposal, we may easily make this expression negative in a selected region.

The positive values of the charge density and the negative values of the quasi-density of the energy in some regions characterize spin 1/2 particles. For spin 0 and spin 1 particles we have the opposite results. The energy density is always positive but the charge distribution takes on positive and negative values.

The origin of the problems with the energy density is clearly the author's believe that the quantum field theories for electrons and other fermions can be understood  ``as theories that describe classical field''. For bosons, and in particular for photons, the quantum state can be populated with many particles having the same wave function. This is how the coherent states of photons and the BEC condensates are created and the field operator acquires the classical meaning. For fermions this is not possible because of the Pauli principle. Each quantum state can be occupied only by one particle and the coherent states that could be associated with a classical field are not possible. The Dirac bispinor can be treated only as the wave function of the electron and not as a carrier of some classical reality. The transition to quantum field theory of electrons proceeds by the second quantization of the wave function and not by the quantization of some imagined classical field. The negative values of the quasi-energy demonstrated in our Comment underscore this fact.

Finally, we would like to point out that the analysis of the ``small electron states'' with the use of a specific example, like the Gaussian wave packets that was considered in \cite{seb}, can be carried out without reference to any particular model. The conclusion can be reached using a simple general argument based entirely on dimensional analysis.

The general proof that there is no finite lower limit on the mean square radius $\langle{\bm r}^2\rangle$ starts from the observation that every normalizable solution of the Dirac equation must have some scale parameter $l$ that controls the size of the wave packet in space. This parameter in the Gaussian wave packet studied in \cite{seb} is equal to $\lambdabar_C/n$, where $\lambdabar_C=\hbar/mc$ is the Compton wave length. Therefore, the mean square radius can be expressed in the form,
\begin{align}
\langle{\bm r}^2\rangle=l^2 g\left(\frac{\lambdabar_C^2}{l^2}\right),
\end{align}
where $g$ is a function of the dimensionless parameter. The function $g$ depends on the chosen solution of the Dirac equation. We cannot say yet that $\langle{\bm r}^2\rangle$ goes to zero when $l$ goes to zero because, in principle, the function $g$ may behave as $\lambdabar_C^2/l^2$ when $l\to 0$. This possibility, however, is excluded by the observation that $l\to 0$ in the argument of $g$ is equivalent to $\lambdabar_C\to\infty$, i.e., the particle becomes massless. For a massless particle there is no intrinsic scale parameter and the mean square radius for every solution must be simply proportional to $l^2$, i.e., $\langle{\bm r}^2\rangle={\rm const}\;l^2$. This argument is quite general. It is valid for all solutions of the free Dirac equation.

\end{document}